\documentstyle[preprint,aps]{revtex}
\begin{document}
\draft
\title{Fluctuations and Landau-Devonshire expansion 
for Barium Titanate}

\author{A. I. Sokolov$^1$, A. K. Tagantsev$^2$}

\address{$^1$Saint Petersburg Electrotechnical University, 
St. Petersburg 197376, Russia; \\
e-mail: ais@sokol.usr.etu.spb.ru}

\address{$^2$Laboratoire de C\'eramique, EPFL, 
CH-1015 Lausanne, Switzerland; \\
e-mail: Alexander.Tagantsev@epfl.ch}

\date{February 20, 2002}
\maketitle

\begin{abstract}
The experimentally observed temperature dependence of the 
quartic coefficients in the Landau-Devonshire expansion for 
BaTiO$_3$ is naturally accounted for within a proper fluctuation 
model. It is explained, in particular, why one of the quartic 
coefficients varies with temperature above $T_c$ while the 
second is constant. The tetragonal phase in BaTiO$_3$ is argued 
to exist essentially due to the thermal fluctuations, while the 
true Landau-Devonshire expansion with temperature-independent 
coefficients favours the rhombohedral ferroelectric phase.
Certain conclusions concerning the temperature dependence of 
the sextic Landau-Devonshire coefficients are also made.  

\vspace{5mm}
\end{abstract}


\eject

\vspace{4mm}

The phase diagram of the barium titanate contains three lines 
of phase transitions and its structure is known to be properly 
reproduced by the phenomenological Landau-Devonshire theory 
\cite{D49}. It is accepted within the phenomenological approach 
that all the coefficients in the Landau free energy expansion 
should be either constant or weak functions of temperature, 
pressure, etc., apart from the quadratic term that changes sign 
crossing the second-order transition line or the low-temperature 
spinodal. For the barium titanate, however, 
the matching with the theory of the experimentally observed 
temperature dependences of the nonlinear dielectric susceptibility 
and spontaneous polarization forces one to allow for the strong 
temperature dependence of some higher-order coefficients 
\cite{DLY,HY,BCG}. Being in conflict with the spirit of the Landau 
theory itself, rapid temperature variation of the quartic and sextic 
coefficients is also quite unexpected from the microscopic point 
of view. Indeed, the barium titanate is a displacive ferroelectric 
with weak anharmonicity that can only results in rather slow 
temperature dependence of the macroscopic parameters 
\cite{VGV}. Moreover, there is an extra point showing how 
unsatisfactory the real situation is: because of the fast 
temperature variation of the certain coefficient, the six-order 
Landau-Devonshire expansion for BaTiO$_3$ with updraded 
coefficients \cite{BC,B} turns out to loose its global stability at 
$T$ = 443 K, i. e. at the temperature that exceeds the 
cubic-tetragonal transition point $T_c$ by only 50 K.   

In this Letter, we will show that the temperature dependence 
of the quartic Landau-Devonshire coefficients, experimentally 
observed in a paraelectric phase of BaTiO$_3$, can be attributed 
to the thermal fluctuations of polarization and naturally 
accounted for within a proper fluctuation model. 
It will be explained, in particular, why in experiments one of 
the quartic coefficients demonstrates well-pronounced variation 
with temperature while the second is temperature independent. 
The conjecture will be put forward concerning the structure of 
the true Landau-Devonshire form for BaTiO$_3$ with the 
constant (temperature independent) coefficients, and the  
first-order fluctuation corrections to the sextic Landau-Devonshire 
coefficients for $T > T_c$ will be found.  

Barium titanate is a displacive ferroelectric undergoing the 
first-order ferroelectric phase transitions, with the six-order 
anharmonicity and electrostriction playing an essential role in 
forming its phase diagram. The electrostriction is known to 
convert the second-order transition, appropriate to the 
clamped crystal, into the first-order one. Strong dipole-dipole 
interaction affects the vibrational spectrum of BaTiO$_3$, 
resulting in a big gap between transversal (soft) and 
longitudinal polarization modes. These main features are 
properly described by the effective Hamiltonian that is the 
natural generalization of the Landau-Devonshire free-energy 
expansion: 
\begin{eqnarray}
\label{eq:H}
H & = & {\frac{1}{2}} \sum_{q} 
\Biggl[\Biggl({\frac{T - T_{0}}{C \epsilon_{0}}} 
+ s^2 q^2\Biggr) {\delta_{\alpha \beta}} 
+ \Delta^2 n_{\alpha} n_ {\beta}\Biggr] 
\phi_{\alpha q} \phi_{\beta q}^{*}  
+ {\frac{1}{2}} \int d^3x \biggl[c_{11} (u_{11}^2 
+ u_{22}^2 + u_{33}^2) 
\nonumber\\
& + & 2 c_{12}(u_{22} u_{33} + u_{33} u_{11} + u_{11} u_{22}) 
+ c_{44}(u_{12}^2 + u_{23}^2 + u_{31}^2)\biggr] 
+  H_{int} + H_{str},
\end{eqnarray}
\begin{eqnarray}
\label{eq:HI}
H_{int} & = & \int d^3x  
\Bigl[ \beta_{1} (\phi_1^4 + \phi_2^4 + \phi_3^4) 
+ 2 \beta_{2} (\phi_1^2 \phi_2^2 + \phi_2^2 \phi_3^2 
+ \phi_3^2 \phi_1^2) 
+ \gamma_{1} (\phi_1^6 + \phi_2^6 + \phi_3^6) 
\nonumber\\
& + & 3 \gamma_{2} (\phi_1^4 \phi_2^2 + \phi_1^2 \phi_2^4 
+ \phi_2^4 \phi_3^2 + \phi_2^2 \phi_3^4 + \phi_3^4 \phi_1^2 
+ \phi_3^2 \phi_1^4) 
+ 6 \gamma_{3} \phi_1^2  \phi_2^2 \phi_3^2 \Bigr],
\end{eqnarray}
\begin{eqnarray}
\label{eq:HS}
H_{str} & = & \int d^3x  
\Bigl\{q_{11} (u_{11} \phi_{1}^2 + u_{22} \phi_{2}^2 
+ u_{33} \phi_{3}^2) + q_{12} [u_{11} (\phi_{2}^2 + \phi_{3}^2) 
+ u_{22} (\phi_{3}^2 + \phi_{1}^2) 
\nonumber\\
& + & u_{33} (\phi_{1}^2 +\phi_{2}^2)] 
+ q_{44} (u_{23} \phi_2 \phi_3 + u_{31} \phi_1 \phi_3 
+ u_{12} \phi_2 \phi_1)\Bigr\}. 
\end{eqnarray}
Here $\phi_{\alpha}$ and $u_{\alpha \beta}$ are the Cartesian 
components of the fluctuating polarization and strain, 
$\phi_{\alpha q}$ stands for the Fourier transform of 
$\phi_{\alpha}(x)$, $n_{\alpha} = {\frac{q_{\alpha}}{q}}$, $C$  
is the Curie constant, 
$\Delta^2 \sim \epsilon_{0}^{-1} \sim s a^{-1}$ 
is the dipole gap in the fluctuation spectrum, $a$ being 
a lattice constant. Contrary to the 
original Landau-Devonshire expansion, the Hamiltonian (1-3)
accounts for the inhomogeneous fluctuations of the polarization 
and elastic strains making it possible to explore the fluctuation 
effects in BaTiO$_3$.  

Dealing with the first-order phase transition, we are in a position 
to study the fluctuation effects in the region where thermal 
fluctuations of the order parameter are weak. 
Hence, in what follows we limit ourselves by the calculation of 
the first-order fluctuation corrections to the quantities of interest. 
The quantities to be found are the full four-leg and six-leg vertices, 
reducing, under zero external momenta, to the effective 
("dressed") quartic $B_i$ and sextic $\Gamma_i$ coefficients, 
entering the Landau-Devonshire expansion. Among five Feynman  
diagrams representing the lowest-order non-trivial terms 
in the perturbative expansions of the vertices mentioned, only 
two one-loop graphs give contributions that rapidly, 
as $(T - T_0)^{-1/2}$, grow up approaching the transition point. 
These graphs have an obvious structure and may be calculated 
in a standart way using the propagator   
\begin{equation}
\label{eq:G}
G_{\alpha \beta}(q) = {\frac{k_B T (\delta_{\alpha \beta} 
- n_{\alpha} n_ {\beta})}
{(C \epsilon_{0})^{-1} (T - T_{0}) + s^2 q^2}} ,
\end{equation}
with a longitudinal part fully neglected. The results for the 
quartic couplings $B_1$ and $B_2$ are found to be: 
\begin{eqnarray}
\label{eq:CC}
B_1 & = & \beta_1 - {\frac{k_B T \sqrt{C \epsilon_0}}
{10 \pi s^3 \sqrt{T - T_0}}} 
(24 \beta_{1}^2 + 4 \beta_{1} \beta_{2} + 6 \beta_{2}^2), 
\nonumber\\
B_2 & = & \beta_2 - {\frac{k_B T \sqrt{C \epsilon_0}}
{10 \pi s^3 \sqrt{T - T_0}}} 
(3 \beta_{1}^2 + 18 \beta_{1} \beta_{2} + 13 \beta_{2}^2), 
\end{eqnarray}
The polynomials in brackets are easily seen to coincide 
with those of the one-loop contributions to the 
renormalization-group $\beta$-functions of the cubic 
ferroelectric \cite{S75,ST79}. It is not surprising since, 
in fact, the same integrals and tensor convolutions are 
evaluated in both cases. The fluctuation correction to $B_1$ 
given by the first equation (5) is consistent with the results 
of Vaks \cite{VGV,V70}, who first evaluated this correction 
and showed that it is essential for the case of BaTiO$_3$.

To proceed further, we have to estimate the "bare" coupling 
constants $\beta_{1}$ and $\beta_{2}$ for BaTiO$_3$ trusting 
upon the experimental data available. Aiming to extract the
necessary information from experiments, one should realize 
that  i) what is measured are not bare but dressed
couplings with the fluctuation contributions included, and  
ii) in experiments, the Landau-Devonshire coefficients are 
measured for stress-free (not clamped) crystals. Hence, in 
order to estimate $\beta_{1}$ and $\beta_{2}$, we have first 
to express them via their analogs for free crystal, 
$\beta_{1}^{f}$ and $\beta_{2}^{f}$. This problem is solved 
by evaluation of the elastic strains caused by the non-zero 
uniform polarization and consequent renormalization of 
coefficients in the relevant Landau-Devonshire expansion 
\cite{D49}. All the elastic and electrostrictive moduli are 
known for BaTiO$_3$ \cite{B}, making corresponding calculations 
straightforward. The final result is as follows: 
\begin{equation}
\label{eq:R}
\beta_{1} = \beta_{1}^{f} + 7.4\cdot10^8 V m^5 C^{-3}, 
\quad \quad
\beta_{2} = \beta_{2}^{f} - 2.3\cdot10^8 V m^5 C^{-3}, 
\end{equation}
Since in the vicinity of $T_c$ the elastic and 
electrostrictive moduli weakly depend on temperature, 
similar relations should be valid for the fluctuation modified 
(dressed) quartic coefficients $B_{1}$, $B_{2}$,  
$B_{1}^{f}$, and $B_{2}^{f}$. 

At the transition point ($T_c$ = 393 K), 
$B_{1}^{f} = -2.0\cdot10^8 V m^5 C^{-3}$,  
$B_{2}^{f} = 1.6\cdot10^8 V m^5 C^{-3}$ \cite{BC,B}
and, hence, $B_{1} = 5.4\cdot10^8 V m^5 C^{-3}$, 
$B_{2} = -0.7\cdot10^8 V m^5 C^{-3}$. It is easy to see 
that for $B_{1} > B_{2}$ the rhombohedral phase 
has lower free energy than the tetragonal one, 
provided the six-order form is isotropic, i. e. does not 
influence their competition. Hence, in the clamped 
crystal the quartic form of the Landau-Devonshire 
expansion strongly favours the phase transition into 
the rhombohedral phase. The same is true for higher 
temperatures $T$ = 415 K and $T$ = 423 K, where 
$B_{1}^{f} = -1.3\cdot10^8 V m^5 C^{-3}$ \cite{DLY},
$B_{1} = 6.1\cdot10^8 V m^5 C^{-3}$ 
and 
$B_{1}^{f} = -1.0\cdot10^8 V m^5 C^{-3}$ \cite{DLY},
$B_{1} = 6.4\cdot10^8 V m^5 C^{-3}$, 
respectively, with $B_{2}^{f}$ and $B_{2}$ kept 
unchanged.

Now we are ready to estimate to what extend the 
fluctuations can modify the behaviour of barium 
titanate in the vicinity of $T_c$.    
As is seen from Eqs.(\ref{eq:CC}), the quartic 
Landau-Devonshire coefficients should vary with 
temperature in a similar way, provided the fluctuation 
corrections to them are of the same order of magnitude. 
Let us compare the magnitudes of the fluctuation terms 
$B_{2}^{(1)}$ and $B_{1}^{(1)}$ within the domain where 
parameters $\beta_{1}$ and $\beta_{2}$ have relevant 
values and signs. The quantity characterizing relative   
weights of $B_{2}^{(1)}$ and $B_{1}^{(1)}$ is their ratio, 
that can be found directly from Eqs. (\ref{eq:CC}). The plot 
of the ratio $R^{(1)} = {\frac{B_{2}^{(1)}}{B_{1}^{(1)}}}$ 
as a function of $r = {\frac{\beta_{1}}{\beta_{2}}}$ is 
shown in Fig. 1. Analyzing this function, one can find, in 
particular, that 
$\bigg|{\frac{B_{2}^{(1)}}{B_{1}^{(1)}}}\bigg|$ does not 
exceed 0.1 if ${\frac{\beta_{1}}{\beta_{2}}}$ lies between
$-$28 and $-$2.3.  

Let us estimate further ${\frac{\beta_{1}}{\beta_{2}}}$  
in BaTiO$_3$. Since, 
according to experiments, $B_{1}$ varies with the 
temperature appreciably, an estimate for $\beta_{1}$ is 
expected to have somewhat limited accuracy. To keep 
the perturbation theory more or less meaningful, one 
should adopt that the fluctuation term $B_{1}^{(1)}$ is, 
at least, two times smaller than the value of $B_{1}$ at 
$T = T_c$. At the transition temperature $B_{1} = 
5.4\cdot10^8 V m^5 C^{-3}$ and this coefficient was 
shown to increase when the temperature grows up. 
Hence, the limitation 
$\beta_{1} \leq {10^9} V m^5 C^{-3}$ 
looks quite reasonable. Another quartic coefficient,
$B_{2}$, does not depend on $T$ and the estimate 
$\beta_{2} = B_{2} = -0.7\cdot10^8 V m^5 C^{-3}$ 
may be considered as an accurate one. 
Since, in any case, $\beta_{1} > B_{1}(T = 393 K) 
= 5.4\cdot{10^8} V m^5 C^{-3}$,  we see that 
$-14 < {\frac{\beta_{1}}{\beta_{2}}} < -8$ in barium titanate. 
It implies that, according to Fig. 1, 
$B_{2}^{(1)} < 0.08 B_{1}^{(1)}$, and the fluctuation 
correction $B_{2}^{(1)}$ is inevitably very small. 
This means that $B_{2}$ practically does 
not depend on $T$ and explanes why in BaTiO$_3$ 
the Landau-Devonshire coefficient $B_{2}^{f}$ 
is insensitive to the temperature. 

Apart from the smallness of $B_{2}^{(1)}$, the theory 
naturally accounts for the experimentally observed 
sign of the fluctuation contribution to $B_{1}$. 
Indeed, as seen from Eqs. (\ref{eq:CC}), a sign of 
$B_{1}^{(1)}$ is completely controlled by the first term 
proportional to $\beta_{1}^2$: big number 24 provides 
a positiveness of the polynomial in brackets avoiding 
any possibility for the rest terms to compete with the 
first one under any values of $\beta_1$ and $\beta_2$. 
Hence, the theory definitely predicts that the 
fluctuation correction $B_{1}^{(1)}$ is negative and, 
therefore, approaching $T_c$ from above $B_{1}$ 
and $B_{1}^{f}$ should vary downward. This 
conclusion is in agreement with experiments 
\cite{DLY,HY}. 

As we have already seen, the experimental data available 
do not allow to fix the true, temperature independent value 
of the Landau-Devonshire coefficient $\beta_{1}$. 
However, there are obvious requirements that enable us 
to improve the crude estimate for $\beta_{1}$ presented 
above. The theory developed, based on the first-order 
perturbative calculations, is believed to account for the 
main features of the behaviour of BaTiO$_3$. Hence, it 
should explain the variation of $B_{1}^{f}(T)$ by, at least, 
${10^8}V m^5 C^{-3}$ \cite{DLY} and remain valid, at worst, 
at the semiquantitative level. It is possible to meet both 
requirements only accepting that the true value 
of $\beta_{1}$ is appreciably bigger than 
$6.4\cdot{10^8}V m^5 C^{-3}$ and appreciably smaller 
than $1.1\cdot{10^9}V m^5 C^{-3}$, i. e. it lies somewhere 
between $7\cdot{10^8}V m^5 C^{-3}$ and 
${10^9} V m^5 C^{-3}$. As a result, the stress-free crystal 
should possess the Landau-Devonshire coefficient that 
obeys the inequalities $-0.4\cdot{10^8}V m^5 C^{-3} < 
\beta_{1}^{f} < 2.6\cdot{10^8}V m^5 C^{-3}$. 

The upper part of this interval is of particular interest. 
The point is that, whenever $\beta_{1}^{f}$ exceeds 
$1.6\cdot{10^8}V m^5 C^{-3} = \beta_{2}^{f}$, the 
crystal with the fluctuations being "switched off" would 
undergo a phase transition into the rhombohedral phase 
rather than into the tetragonal one. If this were true, 
i. e. the inequality $\beta_{1}^{f} > \beta_{2}^{f}$ took 
place, the tetragonal and orthorhombic phases in barium 
titanate would exist essentially due to the thermal 
fluctuations, while the rhombohedral phase would survive 
only at sufficiently low temperatures where the 
fluctuations are weak enough. Thus, the analysis based 
upon the effective Hamiltonian (1) turns out to support 
the conjecture about the fluctuation stabilization of 
the high-temperature ferroelectric phases in BaTiO$_3$ 
formulated first within the microscopic theory \cite{VGV}.    

This conjecture looks rather attractive. It is worthy 
to discuss it in more detail. It turns out that, apart from  
the results the of first-order calculations, there exist two 
extra arguments in favour of the scenario just described. 
The first is as follows. As we have already seen, the 
lowest-order correction to $\beta_{1}^{f}$ is considerable 
and, therefore, the higher-order fluctuation contributions 
can influence the results appreciably. The leading 
perturbative term shifting the first-order estimates is the 
second-order one and it has a sign opposite to that of 
$B_{1}^{(1)}$. One can show that account for the
positive second-order fluctuation term in the course of 
fit of the experimental data \cite{DLY} results in the 
shift of the estimated value of $\beta_{1}^{f}$ upward. 
Hence, within the refined theory the fluctuation origin 
of the tetragonal and orthorhombic phases in BaTiO$_3$ 
would become more plausible.  

The second argument deals with the fluctuation corrections 
to the six-order coefficients in the Landau-Devonshire 
expansion. Let us calculate them. Evaluating the six-leg 
one-loop diagram, we arrive to the following expressions 
for $\Gamma_i$:    
 \begin{eqnarray}
\label{eq:CCC}
\Gamma_{1} & = & \gamma_{1} 
- f(T)(120 \beta_{1} \gamma_{1}
 + 6 \beta_{1} \gamma_{2} 
+ 10 \beta_{2} \gamma_{1} + 18 \beta_{2} \gamma_{2}), 
\nonumber\\
\Gamma_{2} & = & \gamma_{2} 
- f(T)(5 \beta_{1} \gamma_{1} 
+ 63 \beta_{1} \gamma_{2} + 2 \beta_{1} \gamma_{3} 
+ 15 \beta_{2} \gamma_{1} + 63 \beta_{2} \gamma_{2}
+ 6 \beta_{2} \gamma_{3}),
\nonumber\\
\Gamma_{3} & = & \gamma_{3} 
- f(T)(18 \beta_{1} \gamma_{2} 
+ 24 \beta_{1} \gamma_{3} + 54 \beta_{2} \gamma_{2}
+ 58 \beta_{2} \gamma_{3}), 
\end{eqnarray}
where $f(T) = k_B T \sqrt{C \epsilon_0}
\Bigl(10 \pi s^3 \sqrt{T - T_0}\Bigr)^{-1}$.

Above formulas are written down for the physical 
case when the polarization vector has three 
Cartesian components. In fact, the fluctuation 
terms were calculated for the generic model with 
the n-vector order parameter. It was done in order to 
reserve an opportunity for independent check of the 
results obtained. Indeed, as is well known, the model 
with cubic anisotropy possesses the special symmetry 
property under $n = 2$: if one turns the field 
$(\phi_1, \phi_2)$ by $45^o$ in its two-dimensional 
space, the coupling constants are transformed, but the 
structure of the Hamiltonian remains unchanged. This 
implies some exact symmetry relations between 
coupling constants \cite{PS01}, that interrelate dressed 
($B_{i}$, $\Gamma_{i}$) and bare
($\beta_{i}$, $\gamma_{i}$) couplings and can be used 
to approve (or disregard) the results of perturbative 
calculations in the arbitrary order. Having made such a 
check up, we found that the n-vector analog of 
Eqs.(\ref{eq:CCC}) do obey these symmetry relations 
when $n = 2$.

To estimate the magnitudes of fluctuation corrections 
$\Gamma_{i}^{(1)}$, the experimental information about 
$\gamma_{i}$ is necessary. The 
experiments, however, were carried out in the ordered 
(tetragonal) phase and yielded, in particular, strong  
dependence of $\Gamma_{1}$ on temperature.  This 
temperature dependence was already mentioned to be 
so dramatic that, being extrapolated to the paraelectric 
region, makes the Landau-Devonshire form unstable for 
$T > T_c + 50 K$. It is hardly believed therefore that the 
experimental data available can be used to extract 
more or less reliable estimates for $\gamma_{i}$.  

In such a situation it is natural to analyze the general 
structure of the correction terms $\Gamma_{i}^{(1)}$ in 
Eqs. (\ref{eq:CCC}), aiming to find some conclusions 
that are insensitive to the concrete values of 
$\gamma_{i}$. Let us proceed, accepting that 
$\gamma_{1} \sim \gamma_{2} \sim \gamma_{3}$.  
As we have already found, in barium titanate $|\beta_{2}| 
<< |\beta_{1}|$. It means that the magnitudes of 
$\Gamma_{i}^{(1)}$ are determined, in fact, by the 
terms in Eqs. (\ref{eq:CCC}) containing $\beta_{1}$.
The numerical coefficients before these terms are seen 
to be markedly different. The biggest one (120) stands 
in the expression for $\Gamma_{1}^{(1)}$, making 
$\Gamma_{1}$ stronger dependent on temperature than
two other coefficients. Moreover, the structure of 
$\Gamma_{1}^{(1)}$ fixes its sign. Since $\beta_{1} > 0$, 
$\Gamma_{1}^{(1)}$ is negative and the fluctuations 
diminish the Landau-Devonshire coefficient 
$\Gamma_{1}$ provided $\gamma_{1} > 0$. The 
positiveness of $\gamma_{1}$ is, in its turn, inevitable, 
because this coefficient is responsible for the global 
stability of the system outside the fluctuation region. 

So, we see that approaching $T_c$ the coefficient 
$\Gamma_{1}$ decreases more rapidly than 
$\Gamma_{2}$ and $\Gamma_{3}$. On the other hand, 
as one can see, the smaller $\Gamma_{1}$, the more 
stable is the tetragonal phase. Hence, the fluctuations
modify the six-order form in the Landau-Devonshire 
expansion in a way that favours the transition into the 
tetragonal phase. It confirms the conjecture about the 
fluctuation stabilization of the phases lying between 
the cubic and rhombohedral ones at the phase 
diagram of barium titanate.

To conclude, we have shown that the temperature 
dependence of the quartic Landau-Devonshire 
coefficients in a paraelectric phase of BaTiO$_3$ can 
be explained as the fluctuation effect. The theory 
developed naturally accounts for a signs of the 
temperature variations of $B_1$ and makes it clear 
why the second coefficient, $B_2$, is temperature 
independent. The conjecture was formulated that the 
true Landau-Devonshire form for BaTiO$_3$ with the 
temperature independent coefficients favours the 
transition from cubic into the rhombohedral phase, 
and it is the thermal fluctuations what stabilizes the 
tetragonal and orthorhombic phases and provides the 
space for them at the phase diagram.  

The authors acknowledge the financial support of the Russian 
Foundation for Basic Research under Grant No. 01-02-17048 
(A.I.S.), the Ministry of Education of Russian Federation under 
Grant No. E00-3.2-132 (A.I.S.), and the Swiss National Science 
Foundation (A.K.T.).  A.I.S. has benefitted from the hospitality 
of the Laboratoire de C\'eramique, Ecole Polytechnique 
F\'ed\'eral de Lausanne, where this research was done.

\vspace{3cm}
CAPTION

Fig. 1. The ratio of the first-order fluctuation corrections 
$R^{(1)} = {\frac{B_{2}^{(1)}}{B_{1}^{(1)}}}$ 
as a function of the ratio of bare coupling constants 
$r = {\frac{\beta_{1}}{\beta_{2}}}$.

\end{document}